\documentclass[%
 reprint,
 amsmath,amssymb,
 aps,
 pre,
]{revtex4-1}

\usepackage{CJK}
\usepackage{graphicx}% Include figure files
\usepackage{dcolumn}% Align table columns on decimal point
\usepackage{bm}% bold math
\usepackage{color}
\usepackage{amsmath}
\begin{document}

\begin{CJK*}{UTF8}{} % Use default fonts from CJK (see below)

\preprint{APS/123-QED}

\title{The Meniscus on the Outside of a Circular Cylinder: From Microscopic to Macroscopic Scales}

\author{Yanfei Tang ({\CJKfamily{gbsn}唐雁飞})}
\affiliation{Department of Physics, Center for Soft Matter and Biological Physics,
and Macromolecules Innovation Institute, Virginia Polytechnic Institute and State University,
Blacksburg, Virginia 24061, USA}%Lines break automatically or can be forced with \\
\author{Shengfeng Cheng ({\CJKfamily{gbsn}程胜峰})}
\email{chengsf@vt.edu}
\affiliation{Department of Physics, Center for Soft Matter and Biological Physics,
and Macromolecules Innovation Institute, Virginia Polytechnic Institute and State University,
Blacksburg, Virginia 24061, USA}

\date{\today}% It is always \today, today,

\begin{abstract}
We systematically study the meniscus on the outside of a small circular cylinder vertically immersed in a liquid bath in a cylindrical container that is coaxial with the cylinder. The cylinder has a radius $R$ much smaller than the capillary length, $\kappa^{-1}$, and the container radius, $L$, is varied from a small value comparable to $R$ to $\infty$. In the limit of $L \ll \kappa^{-1}$, we analytically solve the general Young-Laplace equation governing the meniscus profile and show that the meniscus height, $\Delta h$, scales approximately with $R\ln (L/R)$. In the opposite limit where $L \gg \kappa^{-1}$, $\Delta h$ becomes independent of $L$ and scales with $R\ln (\kappa^{-1}/R)$. We implement a numerical scheme to solve the general Young-Laplace equation for an arbitrary $L$ and demonstrate the crossover of the meniscus profile between these two limits. The crossover region has been determined to be roughly $0.4\kappa^{-1} \lesssim L \lesssim 4\kappa^{-1}$. An approximate analytical expression has been found for $\Delta h$, enabling its accurate prediction at any values of $L$ that ranges from microscopic to macroscopic scales.

%\begin{description}
%\end{description}
\end{abstract}

\maketitle

\end{CJK*}

%\tableofcontents

\section{introduction} \label{sec:intro}

A liquid meniscus as a manifestation of capillary action is ubiquitous in nature and our daily life. For example, its formation and motion play critical roles in water uptake in plants \cite{McElrone2013}. Capillary adhesion due to the formation of menisci between solid surfaces makes wet hair to stick together and allows kids to build sandcastles \cite{Hornbaker1997}. Menisci are also involved in many technologies and industrial processes \cite{Yuan2013} such as meniscus lithography \cite{Kang2011}, dip-pen nanolithography \cite{Piner1999}, dip-coating (Langmuir-Blodgett) assembly of nanomaterials \cite{Ghosh2007,Tao2008,Cote2009}, meniscus-mediated surface assembly of particles \cite{Cavallaro2011}, meniscus-assisted solution printing \cite{He2017}, etc.

A meniscus system frequently discussed in the literature is the one formed on the outside of a circular cylinder that is vertically immersed in a liquid bath. One application of this geometry is the fabrication of fiber probes by chemical etching \cite{Takahashi1990}. A cylinder with radius at the nanometer scale has also been attached to the tip of an atomic force microscope to perform nano-/micro-Whilhemy and related liquid property measurements \cite{Yazdanpanah2008}. The shape of the meniscus is governed by the Young-Laplace equation \cite{DeGennes2004}. Extensive studies have been reported for the scenario where the liquid bath is unbound and the lateral span of the liquid-vapor interface is much larger than the capillary length of the liquid \cite{Derjaguin1946, White1965, Huh1969, James1974, Lo1983, Alimov2014}. Different methods have been applied in these studies, including numerical integration \cite{White1965, Huh1969} and analytical approaches such as matched asymptotic expansions \cite{James1974, Lo1983, Alimov2014} and hodograph transformations for cylinders with complex shapes \cite{Alimov2014}. An approximate formula has been derived for the meniscus height, which depends on the radius of the cylinder and the contact angle of the liquid on the cylinder surface \cite{Derjaguin1946,James1974}. The meniscus exerts a force that either drags the cylinder into or expels it from the liquid depending on if the contact angle is acute or obtuse. A recent study of the meniscus rise on a nanofiber showed that the force on the nanofiber highly depends on the lateral size of the liquid-vapor interface if this size is smaller than the capillary length \cite{DeBaubigny2015}. 

\begin{figure}[ht]
  \centering
  \includegraphics[width = 0.45 \textwidth]{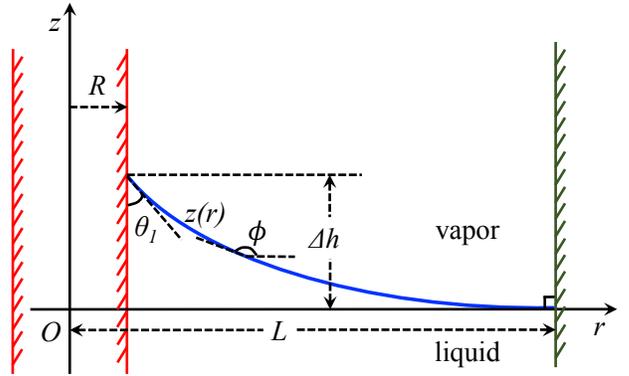}
  \caption{A rising meniscus on the outside of a circular cylinder vertically immersed in a liquid bath confined in a cylindrical container that is coaxial with the cylinder.}
  \label{fg:fig1}
\end{figure}

In this paper we consider a geometry as sketched in Fig.~\ref{fg:fig1} where a small circular cylinder vertically penetrating a liquid bath that is confined in a cylindrical container. With the cylinder and the container being coaxial, the system has axisymmetry that enables certain analytical treatments. By fixing the contact angle on the surface of the container to be $\pi/2$, we have a meniscus that systematically transits from being laterally confined to unbound, when the size of the container is increased. For such a system, the meniscus profile is governed by the general Young-Laplace equation that was first studied by Bashforth and Adams more than a century ago \cite{Bashforth1883}. This equation has been discussed in various systems including liquid in a tube \cite{Concus1968}, sessile and pendant droplets \cite{Obrien1991,Srinivasan2011} and a capillary bridge between two spheres \cite{Mazzone1986}.

In the limit where the size of the cylindrical container is much smaller than the capillary length, the gravitational term in the Young-Laplace equation can be neglected and the equation becomes analytically solvable. Solutions have been reported for various capillary bridges between solid surfaces \cite{Butt2009,Orr1975,Kruyt2017} and tested with molecular dynamics simulations \cite{Cheng2014PRE,Cheng2016Langmuir}. We have obtained a solution for the meniscus in Fig.~\ref{fg:fig1} based on elliptic integrals when the lateral size of the meniscus is small and found that the meniscus height depends on the container size logarithmically. We further numerically solve the full Young-Laplace equation for an arbitrary container size and find that the meniscus height approaches an upper limit found in some early work when the lateral span of the interface is much larger than the capillary length \cite{Derjaguin1946,Huh1969,James1974}. Finally, we find an approximate expression of the meniscus height on the cylinder that is applicable to any lateral size of the liquid-vapor interface. This work is the basis of a related work on the wetting behavior of particles at a liquid-vapor interface \cite{Tang2018sphere}, where the theoretical results presented here are applied to study the detachment of a spherical particle from a liquid bath.

\section{Theoretical Considerations} \label{sec:theory}

\subsection{General Equation of the Meniscus Shape} \label{sec:2a}

The geometry of the system considered in this paper is sketched in Fig. \ref{fg:fig1}. A circular cylinder with radius $R$ is immersed in a liquid bath confined in a cylindrical wall with radius $L>R$. The cylinder and the wall are coaxial and the system is thus axisymmetric. The shape of the meniscus in this ring-shaped tube is determined by the surface tension of the liquid, the contact angles on the two surfaces, and possibly gravity. Our interest is to examine the crossover from the case where $L-R$ is small to the case where the cylinder is immersed in a liquid bath with an infinite lateral span. Since in the latter limit the liquid-vapor interface is flat at locations far away from the cylinder, we will set the contact angle on the wall to be $\pi/2$. Then a meniscus will rise (depress) on the outside of the cylinder if the contact angle on its surface, $\theta_1$, is smaller (larger) than $\pi/2$. The case where $\theta_1=\pi/2$ is trivial with the liquid-vapor interface being flat everywhere. Here we focus on the case with $\theta_1 < \pi/2$, where a meniscus rises on the cylinder and generates a force to pull the cylinder into the liquid bath. However, the final results on predicting the meniscus height also apply to the case where $\theta_1 > \pi/2$.

The equilibrium shape of the meniscus is governed by a form of the Young-Laplace equation studied by Bashforth and Adams before \cite{Bashforth1883},
\begin{align}
\frac{z''}{(1 + z'^2)^{3/2}} + \frac{z'}{r (1 + z'^2)^{1/2}}
= \frac{\Delta p}{\gamma} + \frac{\Delta \rho g z}{\gamma}~,
\label{eq:Young-Laplace}
\end{align}
where $z(r)$ is the meniscus height at distance $r$ from the central axis of the cylinder, $z' \equiv \frac{\mathrm{d}z}{\mathrm{d}r}$, $z'' \equiv \frac{\mathrm{d}^2z}{\mathrm{d}r^2}$, $\Delta p$ is the pressure jump from the vapor to the liquid phase at $r=L$ and $z=0$, $\gamma$ is the surface tension of the liquid, $\Delta \rho \equiv \rho_l - \rho_v$ is the difference of the liquid and vapor densities, and $g$ is the gravitational constant. A brief derivation of this equation is provided in \ref{sec:app1}. In the following discussion, we use a water-air liquid interface at $25^{\circ}$C as an example, for which $\gamma \approx 0.072 \, \mathrm{N/m}$ and $\Delta \rho \approx 10^3 \, \mathrm{kg/m^3}$.

To facilitate discussion, we define $2 \tilde{H} \equiv \frac{\Delta p}{\gamma}$ and $\kappa^2 \equiv \frac{\Delta \rho g}{\gamma}$, i.e., $\kappa^{-1} = \sqrt{\frac{\gamma}{\Delta \rho g}}$ is the so-called capillary length, which is a characteristic length scale of the problem. For water at $25^{\circ}$C, $\kappa^{-1} \approx 2.7$ mm. Eq.~(\ref{eq:Young-Laplace}) can then be made dimensionless via a variable change 
\begin{align}
x \equiv \kappa r~,\quad y \equiv \kappa z~.
\end{align}
The result is the following nonlinear differential equation
\begin{align} \label{eq:YL_dimensionless}
\frac{y''}{(1 + y'^2)^{3/2}} + \frac{y'}{x(1 + y'^2)^{1/2}} 
= \frac{2 \tilde{H}}{\kappa} + y~, 
\end{align}
with boundary conditions
\begin{subequations}  
  \begin{align}
	y^\prime &= -\cot\theta_1 \quad \text{at} \quad x = \kappa R~, \label{eq:BC_dimensionless_a} \\
	y^\prime &= 0 \quad \text{at} \quad x = \kappa L \quad \text{and}  \quad y = 0~. \label{eq:BC_dimensionless_b} 
  \end{align}
\end{subequations}

As pointed out in Ref. \cite{Concus1968}, Eq. (\ref{eq:YL_dimensionless}) is invariant under the transformation $y\rightarrow -y$, $\theta_1 \rightarrow \pi - \theta_1$, and $\tilde{H}\rightarrow-\tilde{H}$, indicating the symmetry between a rising and a depressing meniscus. This second-order nonlinear differential equation can be rewritten in terms of the local tilt angle of the liquid-vapor interface, $\phi$, as defined in Fig.~\ref{fg:fig1}. Since $y^\prime \equiv \frac{\mathrm{d}y}{\mathrm{d}x} = \frac{\mathrm{d}z}{\mathrm{d}r} = \tan \phi$, Eq. (\ref{eq:YL_dimensionless}) then becomes
\begin{align} \label{eq:YL_dimensionless_2}
\frac{\mathrm{d}\sin \phi}{\mathrm{d}x} + \frac{\sin \phi}{x}
= - \frac{2 \tilde{H}}{\kappa} - y~.
\end{align}
Eq. (\ref{eq:YL_dimensionless_2}) and $\frac{\mathrm{d}y}{\mathrm{d}x} = \tan \phi$ can be further rewritten into a pair of coupled first-order nonlinear differential equations in terms of $x(\phi)$ and $y(\phi)$,
\begin{subequations}  
  \begin{align}
	\frac{\mathrm{d}x}{\mathrm{d}\phi} 
	= - (\frac{2\tilde{H}}{\kappa} + y + \frac{\sin \phi}{x})^{-1} 
	\cos \phi~, \\	
	 \frac{\mathrm{d}y}{\mathrm{d}\phi} 
	= - (\frac{2\tilde{H}}{\kappa} + y + \frac{\sin \phi}{x})^{-1} 
	\sin \phi~. 
  \end{align}
  \label{eq:YL_1st_order}
\end{subequations}
with boundary conditions
\begin{subequations}  
  \begin{align}
	\phi &= \phi_1 \quad \text{at} \quad x = \kappa R~, \label{eq:BC_1st_order_a} \\
	\phi &= \phi_2 \quad \text{at} \quad x = \kappa L \quad \text{and}  \quad y = 0~, \label{eq:BC_1st_order_b} 
  \end{align}
\end{subequations}
where $\phi_1 = \theta_1 + \pi/2$ and $\phi_2 = \pi$ for the system sketched in Fig.~\ref{fg:fig1}. Here $\theta_2$ is the contact angle on the wall and is fixed at $\pi/2$ in this paper. Generally, $\phi_2 = \frac{3\pi}{2} - \theta_2$ for $0\le \theta_2 \le \pi$.

In a general case, Eq.~(\ref{eq:YL_1st_order}) can be numerically solved by the shooting method \cite{Press2007}. For the case where contact angle $\theta_1$ is close to $\pi/2$, a zero-order solution is provided in \ref{sec:app2}. For a general contact angle $\theta_1$, analytical solutions of the meniscus can be found when $L \ll \kappa^{-1}$, where the terms on the right sides of Eqs~(\ref{eq:Young-Laplace}), (\ref{eq:YL_dimensionless}), and (\ref{eq:YL_1st_order}) due to gravity are negligible [Sec.~\ref{sec:2b}]. In the opposite limit where $L \gg \kappa^{-1}$, the $\Delta p$ term is negligible and an approximate solution of the capillary rise on the outside of a small cylinder with $R \ll \kappa^{-1}$ was found before by James using the method of asymptotic matching expansions [Sec.~\ref{sec:2c}]. Below we discuss these limits and numerical solutions of Eq.~(\ref{eq:YL_1st_order}) for $R \ll \kappa^{-1}$ and an arbitrary $L$ (which is of course larger than $R$). The results naturally show the crossover from one limit ($R\ll L \ll \kappa^{-1}$) to the other ($R \ll \kappa^{-1} \ll L$).

\subsection{Analytical Solution in the $L \ll \kappa^{-1} $ Limit} \label{sec:2b}

When the radius of the cylindrical wall is small, i.e., $L \ll \kappa^{-1}$, the Bond number $gL^2\Delta\rho/\gamma \ll 1$. As a result, the gravity's effect can be ignored and Eq.~(\ref{eq:Young-Laplace}) reduces to 
\begin{align}
\frac{z''}{(1 + z'^2)^{3/2}}  + \frac{z'}{r (1 + z'^2)^{1/2}}= 2\tilde{H}~,
\label{eq:YL_no_g}
\end{align}
with $\tilde{H}$ being the local mean curvature of the liquid-vapor interface. This equation has been solved analytically before for a capillary bridge between a sphere and a flat surface \cite{Orr1975, Rubinstein2014}. Here we use the same strategy to solve it for the meniscus in a ring-shaped container as depicted in Fig.~\ref{fg:fig1}.

It is convenient to introduce reduced variables $X = r/R,\ Y=z/R$ and a parameter $u = \sin \phi$. Eq. (\ref{eq:YL_no_g}) is then simplified as
\begin{align}\label{eq:yl_reduce}
-2H = \frac{\mathrm{d}u}{\mathrm{d}X} + \frac{u}{X}~,
\end{align}
where $H$ is the dimensionless mean curvature defined as $H \equiv R \tilde{H}$. The boundary conditions are
\begin{subequations}  
  \begin{align}
	\phi &= \phi_1 \quad \text{at} \quad X = 1~, \label{eq:bc3a} \\
	\phi &= \phi_2 \quad \text{at} \quad X = l \quad \text{and} \quad Y = 0~, \label{eq:bc3b} 
  \end{align}
\end{subequations}
where $\phi_1 = \theta_1 + \pi/2$, $\phi_2 = \pi$, and $l = L/R$ is the scaled radius of the cylindrical container. The solution for Eq. (\ref{eq:yl_reduce}) is 
\begin{align}\label{eq:solution}
u = \frac{c}{4H X} - H X~.
\end{align}
The boundary condition in Eq.~(\ref{eq:bc3b}) yields $c = 4H^2l^2$. The other boundary condition in Eq.~(\ref{eq:bc3a}) can then be used to determine the dimensionless mean curvature as
\begin{align}\label{eq:H}
H = \frac{\sin \phi_1}{l^2 - 1}~.
\end{align}

\begin{figure}[ht]
  \centering
  \includegraphics[width = 0.45 \textwidth]{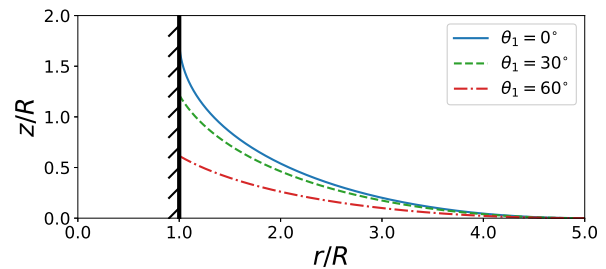}
  \caption{Meniscus profiles from the analytic solution in Eqs.~(\ref{eq:xphi}) and (\ref{eq:yphi}) for $L/R = 5$ and $\theta_1 = 0^{\circ}$ (blue solid line), $30^{\circ}$ (green dashed line), and $60^{\circ}$ (red dash-dotted line).}
  \label{fg:fig2}
\end{figure}

From Eq. (\ref{eq:solution}) and $\mathrm{d}Y/\mathrm{d}X = \tan \phi$, we obtain the analytic solution of the meniscus profile,
\begin{align}
X(\phi) &= \frac{1}{2H}(-\sin \phi + \sqrt{\sin^2 \phi + c})~,   \label{eq:xphi} \\
Y(\phi) &= \frac{1}{2H} \int_{\phi_2}^{\phi} (-\sin t + 
\frac{\sin^2 t}{\sqrt{\sin^2 t +c}})\,\mathrm{d} t~. \label{eq:yphi}
\end{align}
The solution for $Y(\phi)$ in Eq.~(\ref{eq:yphi}) can be written in terms of elliptic integrals,
\begin{align}
Y(\phi) &=& \frac{1}{2H}(\cos \phi - \cos \phi_2)
+ \frac{\sqrt{c}}{2H} \Big[ E(\phi,j) \nonumber \\ 
& & - E(\phi_2,j)  - F(\phi,j) + F(\phi_2, j) \Big]~, \label{eq:yphi2}
\end{align}
where $j^2 \equiv -\frac{1}{c}$, $E(\phi, j) = \int_0^{\phi} \sqrt{1 - j^2 \sin^2 t} \, \mathrm{d}t$ is the incomplete elliptic integral of the second kind, and $F(\phi, j) = \int_0^{\phi} \frac{1}{\sqrt{1 - j^2 \sin^2 t}}\, \mathrm{d}t$ is the incomplete elliptic integral of the first kind. The meniscus rise can be easily computed as $\Delta h = R Y(\phi_1)$, or explicitly,
\begin{align}\label{eq:height_elliptic}
\Delta h &= & \frac{R}{2H}(1 - \sin \theta_1) + \frac{R \sqrt{c}}{2H} 
\Big[ F(\pi/2 - \theta_1, j) \nonumber \\
& & - E(\pi/2 - \theta_1, j)\Big]~.
\end{align}
Some examples of the meniscus profile are shown in Fig.~\ref{fg:fig2} for $L/R = 5$ and $\theta_1 = 0^\circ$, $30^\circ$, and $60^\circ$, respectively.

The analytical prediction in Eq.~(\ref{eq:height_elliptic}) actually indicates that $\Delta h \sim R\ln(L/R)$ when $\kappa^{-1} \gg L > R$. To see this scaling behavior transparently, we examine the limit where $\kappa^{-1} \gg L \gg R$, i.e., the cylinder is much smaller than the cylindrical container and both are much smaller than the capillary length. In this limit we can take $l \gg 1$ and $j^2 \rightarrow -\infty$, and approximate the elliptic integrals in Eq.~(\ref{eq:height_elliptic}) by series expansions. The mathematical derivation is provided in \ref{sec:app3}. The final result on the meniscus height is
\begin{align} \label{eq:height_logL_2}
\Delta h = R \cos \theta_1 \Big[  \ln \frac{2L}{ R(1 + \sin \theta_1)} - \frac{1}{2} \Big]~.
\end{align}

A more intuitive way to see the logarithmic behavior is to note that in the limit of $l \gg 1$, the dimensionless mean curvature $H$ approaches zero and Eq. (\ref{eq:Young-Laplace}) can be rewritten as \cite{DeGennes2004,DeBaubigny2015}, 
\begin{align}\label{eq:yl0}
\frac{r}{(1 + r'^2)^{1/2}} = R \cos \theta_1~, 
\end{align}
where $r' \equiv \frac{\mathrm{d}r}{\mathrm{d}z}$. The solution of this equation is known as a catenary curve \cite{DeGennes2004}. The meniscus is thus a catenoid with its generatrix given by
\begin{align}
z(r) = R \cos \theta_1 \ln \Big[
\frac{L + (L^2-R^2 \cos^2 \theta_1)^{1/2}}{r + (r^2-R^2 \cos^2 \theta_1)^{1/2}}
\Big]~.
\end{align}
The meniscus height can be computed as $\Delta h = z(R)$ and an approximate expression is 
\begin{align}
\Delta h = R \cos \theta_1 \ln \Big[ \frac{2L}{R(1 + \sin \theta_1)} \Big]~,
\label{eq:height_logL}
\end{align}
where the condition $L/R \gg 1 \ge \cos \theta_1$ is used. In both Eqs.~(\ref{eq:height_logL_2}) and (\ref{eq:height_logL}) the scaling dependence of $\Delta h$ on $R\ln (L/R)$ is obvious. However, the expression in Eq.~(\ref{eq:height_logL_2}) for $\Delta h$ is smaller than Eq.~(\ref{eq:height_logL}) by $(R \cos \theta_1 )/2$. This difference stems from the different boundary conditions at the wall. Eq.~(\ref{eq:height_logL_2}) is based on Eq.~(\ref{eq:yphi}) which describes a meniscus that meets the wall with a contact angle $\pi/2$. However, Eq.~(\ref{eq:height_logL}) is based on a catenary curve, for which the contact angle at the wall is close to but not exactly $\pi/2$.

\subsection{Approximate Solution in the $L \gg \kappa^{-1}$ Limit} \label{sec:2c}

In the literature, the meniscus on the outside of a circular cylinder vertically penetrating a liquid bath was mostly investigated for the case where the lateral span of the liquid bath is much larger than the capillary length \cite{Derjaguin1946, White1965, Huh1969, James1974, Lo1983, Alimov2014}, i.e., $L \gg \kappa^{-1}$. In this limit, $\tilde{H}\rightarrow 0$ and the Young-Laplace equation that needs to be solved reads
\begin{align} \label{eq:yl_g}
\frac{y''}{(1 + y'^2)^{3/2}} + \frac{y'}{x(1 + y'^2)^{1/2}} 
= y~.
\end{align}
The boundary condition Eq. (\ref{eq:BC_dimensionless_a}) remains the same but Eq. (\ref{eq:BC_dimensionless_b}) is replaced by
\begin{align}
y^\prime &= 0 \quad \text{at} \quad x \rightarrow \infty \quad \text{and} \quad y = 0~. \label{eq:bc4b} 
\end{align}
\begin{figure}[htb]
  \centering
  \includegraphics[width = 0.45 \textwidth]{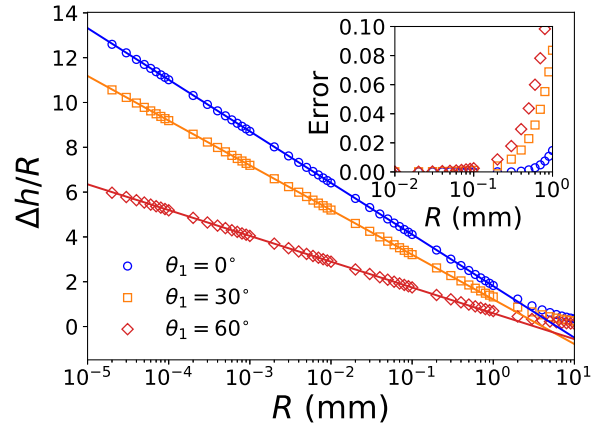}
    \caption{Comparison of the meniscus height ($\Delta h$) between the Derjaguin-James formula (Eq.~(\ref{eq:derja}), solid lines) and numerical results (symbols) using Huh-Scriven's integration scheme \cite{Huh1969} as a function of the radius of the cylinder, $R$, for different contact angles: $\theta_1= 0^{\circ}$ (blue line and {\scriptsize $\bigcirc$}), $30^{\circ}$ (orange line and {\scriptsize $\square$}), and $60^{\circ}$ (red line and {\large $\diamond$}). The lateral span of the liquid bath is treated as infinite by using Eq.~(\ref{eq:bc4b}) as a boundary condition. Inset: the relative deviation of the numerical results on $\Delta h$ from the prediction based on the Derjaguin-James formula is plotted against $R$.
    }
  	\label{fg:fig3}
\end{figure}

Eq.~(\ref{eq:yl_g}) has been studied with methods of numerical integration \cite{White1965, Huh1969} and matched asymptotic expansions \cite{James1974, Lo1983}. The meniscus height is approximately given by the Derjaguin-James formula \cite{Derjaguin1946, James1974},
\begin{align}
\Delta h = R \cos \theta_1 
\Big[\ln \frac{4\kappa^{-1}}{R \left(1 + \sin \theta_1\right)} - E \Big]~, \label{eq:derja}
\end{align}
where $E = 0.57721...$ is the Euler-Mascheroni constant. Eq.~(\ref{eq:derja}) is expected to predict the meniscus height accurately when the radius of the cylinder is much smaller than the capillary length that is in turn much smaller than the lateral span of the liquid bath, namely $R \ll \kappa^{-1} \ll L$. A comparison between the Derjaguin-James formula and numerical results has been fully discussed in Ref. \cite{James1974} for $L\rightarrow \infty$. This comparison is revisited in Fig.~\ref{fg:fig3}. Practically, for water at room temperature it is legitimate to use the Derjaguin-James formula to estimate the meniscus height on a cylinder when its radius is less than about $0.1$ mm.

\section{Numerical Results and Discussion} \label{sec:result}

As discussed in Sec.~\ref{sec:2a}, the general Young-Laplace equation [Eq.~(\ref{eq:YL_dimensionless})] can only be solved numerically. We rewrite Eq.~(\ref{eq:YL_dimensionless}) into a pair of coupled firs-order differential equations [Eq.~(\ref{eq:YL_1st_order})] and adopt the shooting method to obtain their numerical solutions for a given $R$ that is much smaller than $\kappa^{-1}$ and an arbitrary $L$ that varies from $2R$ to a value much larger than $\kappa^{-1}$.

\begin{figure*}[ht]
  \centering
  \includegraphics[width = 1.0 \textwidth]{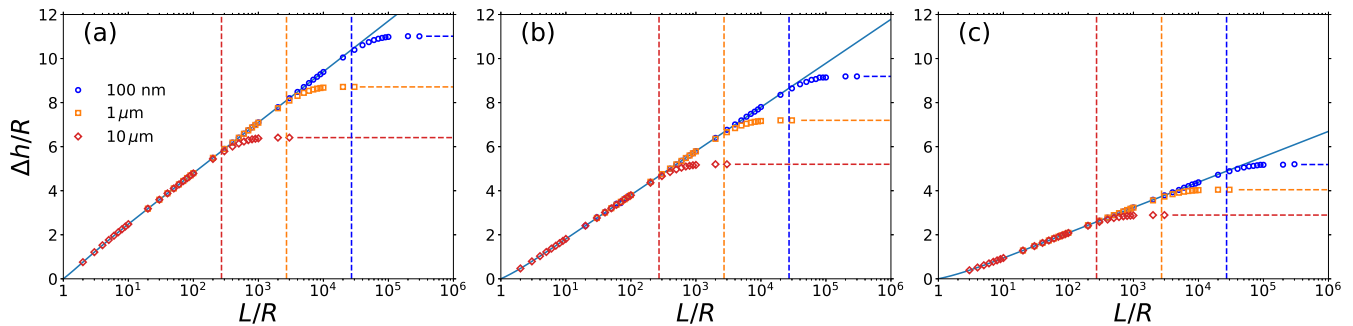}
    \caption{
The meniscus height, $\Delta h$, for different contact angles on the surface of the cylinder: (a) $\theta_1 = $ $0^{\circ}$, (b) $30^{\circ}$, and (c) $60^{\circ}$ as a function of the lateral span of the liquid bath, $L$. The solid line is the analytical expression of $\Delta h$ for $L \ll \kappa^{-1}$ [Eq.~(\ref{eq:height_elliptic})]. The horizontal dashed lines are the predictions of the Derjaguin-James formula for $L \gg \kappa^{-1}$ [Eq.~(\ref{eq:derja})]. The vertical horizontal lines indicate where $L = \kappa^{-1}$. The symbols are numerical solutions of Eq.~(\ref{eq:YL_1st_order}) for an arbitrary $L$ using the shooting method. Data are for cylinders with different radii: $R = 100 \, \mathrm{nm}$ (red {\scriptsize $\bigcirc$}), $1 \, \mu \mathrm{m}$ (orange {\scriptsize $\square$}), and $10 \, \mu \mathrm{m}$ (blue {\large $\diamond$}). Both $\Delta h$ and $L$ are normalized by $R$.
    }
	\label{fg:fig4}
\end{figure*} 

Figure~\ref{fg:fig4} shows numerical solutions of the meniscus height on a circular cylinder immersed vertically in water when $L$ is varied. Cylinders with radii $R$ from 100 nm to 10 $\mu$m and contact angles $\theta_1$ from 0 to $60^\circ$ are used as examples. The data show the following trends. When $L$ is smaller than 1 mm, i.e., $L/R < 10^2$ for $R = 10 \, \mu \mathrm{m}$, $L/R < 10^3$ for $R = 1 \, \mu \mathrm{m}$, and $L/R < 10^4$ for $R = 100 \, \mathrm{nm}$, the meniscus height $\Delta h$ is well predicted by Eq.~(\ref{eq:height_elliptic}), which is derived with gravity ignored. In this limit, $\Delta h$ grows with $L$ logarithmically. In the other limit where $L$ is larger than 10 mm, i.e., $L/R > 10^3$ for $R = 10 \, \mu \mathrm{m}$, $L/R > 10^4$ for $R = 1 \, \mu \mathrm{m}$, and $L/R > 10^5$ for $R = 100 \, \mathrm{nm}$, the meniscus height fits to the Derjaguin-James formula in Eq.~(\ref{eq:derja}), which is derived assuming $R \ll \kappa^{-1}$ and $L \rightarrow \infty$. For $L$ with an intermediate value between 1 mm and 10 mm, the numerical data on the meniscus rise show clearly the crossover between the logarithmic regime [Eq.~(\ref{eq:height_elliptic})] and the saturation regime described by the Derjaguin-James formula. The latter thus provides an upper bound of the meniscus rise on the outside of a circular cylinder with a radius much smaller than the capillary length.

The results in Fig.~\ref{fg:fig4} indicates that for a cylinder with $R \ll \kappa^{-1}$, the Young-Laplace equation without gravity as shown in Eq.~(\ref{eq:YL_no_g}) can be used to describe the meniscus on the outside of the cylinder when $L \lesssim 0.4\kappa^{-1}$, while the liquid bath can be considered as unbounded and the Derjaguin-James formula applies when $L \gtrsim 4\kappa^{-1}$. The range $0.4\kappa^{-1} \lesssim L \lesssim 4\kappa^{-1}$ is the crossover region in which the full Young-Laplace equation [Eqs.~(\ref{eq:Young-Laplace}), (\ref{eq:YL_dimensionless}), (\ref{eq:YL_dimensionless_2}), or (\ref{eq:YL_1st_order})] needs to be employed. This conclusion seems to hold for other liquids with different capillary lengths. For example, we have solved Eq.~(\ref{eq:YL_1st_order}) numerically for a hexadecane-water mixture at 25 $^{\circ}$C, for which $\kappa^{-1} = 4.824$ mm, and found roughly the same crossover zone.

An interesting finding is that the intersection between the solid line from Eq.~(\ref{eq:height_elliptic}) and the corresponding dashed line from the Derjaguin-James formula in Eq.~(\ref{eq:derja}) occurs at $L \approx 1.85 \kappa^{-1}$ for all the systems considered here. The relationship can be understood if we compare Eq.~(\ref{eq:height_logL_2}), which is an approximate form of Eq.~(\ref{eq:height_elliptic}) on the meniscus height in the limit of $\kappa^{-1} \gg L \gg R$, to the Derjaguin-James formula in Eq.~(\ref{eq:derja}). At $L = 2 e^{1/2 -E}\kappa^{-1} \approx 1.85\kappa^{-1}$, the two predictions are equal. This estimate is in perfect alignment with the discovery that at $L \approx 1.85 \kappa^{-1}$, the meniscus height from Eq.~(\ref{eq:height_elliptic}) matches that predicted by the Derjaguin-James formula. In a related work, we find that $L \approx 1.85 \kappa^{-1}$ is also the saturation length of the lateral span of a liquid-vapor interface when discussing how the effective spring constant experienced by a detaching particle depends on the lateral size of the interface \cite{Tang2018sphere}. Note that Eq.~(\ref{eq:height_elliptic}) holds for $\kappa^{-1} \gg L > R$ and is thus more general than Eq.~(\ref{eq:height_logL_2}), which requires $\kappa^{-1} \gg L \gg R$. Our numerical results indicate that Eq.~(\ref{eq:height_elliptic}) provides a good estimate of $\Delta h$ for $L$ up to about $0.4\kappa^{-1}$.

Based on this observation and the finding that the crossover zone, $0.4\kappa^{-1} \lesssim L \lesssim 4\kappa^{-1}$, is relative small, we propose that for a cylinder with radius $R \ll \kappa^{-1}$ and vertically immersed in a liquid bath with lateral span designated as $L$, the meniscus height on the outside of the cylinder can be computed using Eq.~(\ref{eq:height_elliptic}) with the parameter $l$ given as follows,
\begin{equation}
l = \begin{cases}
L/R &\text{if $L \le 1.85\kappa^{-1}$~,}\\
1.85\kappa^{-1}/R &\text{if $L > 1.85\kappa^{-1}$~.}
\end{cases}
\label{eq:height_all}
\end{equation}
Note that the parameter $l$, in addition to $\theta_1$ and $R$, enters in the computation of the parameters $H$, $c$, and $j$ in Eq.~(\ref{eq:height_elliptic}). For $L \le 1.85\kappa^{-1}$, the meniscus height $\Delta h$ depends on $L$ logarithmically while it saturates to the upper bound expressed in the Derjaguin-James formula when $L > 1.85\kappa^{-1}$. Our numerical data indicate that Eq.~(\ref{eq:height_elliptic}) with $l$ from Eq.~(\ref{eq:height_all}) is quite accurate for the meniscus height. Even within the crossover region $0.4\kappa^{-1} \lesssim L \lesssim 4\kappa^{-1}$, the relative deviation of the actual meniscus height from the prediction based on Eqs.~(\ref{eq:height_elliptic}) and (\ref{eq:height_all}) is less than $5\%$, as shown in Fig.~\ref{fg:figS1} in \ref{sec:app4}.

By carefully examining the relative error of using Eqs.~(\ref{eq:height_elliptic}) and (\ref{eq:height_all}) to compute the meniscus height $\Delta h$ and how the error depends on $L$, $R$, and $\theta_1$ [see \ref{sec:app4} for detail], we arrive at an approximate analytical expression of $\Delta h$ for an arbitrary $L$ that reads
\begin{equation}
\Delta h = \Delta h (\textrm{elliptic}) \times \{1-m(\kappa L) [\kappa R (1+\sin \theta_1)]^{0.12}\}~,
\label{eq:height_final}
\end{equation}
where $ \Delta h (\textrm{elliptic})$ is the meniscus height from Eq.~(\ref{eq:height_elliptic}) based on elliptic integrals with the parameter $l$ given in Eq.~(\ref{eq:height_all}) and $m(\kappa L)$ is a universal function given as follows,
\begin{equation}
m(x) = \begin{cases}
0.085 \exp{\left[ (x-1.85)^{1.83}/0.74 \right]} &\text{if $x \le 1.85$~,}\\
0.085 \exp{\left[ (1.85-x)/0.875 \right]} &\text{if $x > 1.85$~.}
\end{cases}
\label{eq:fit_kink_function}
\end{equation}
Note that $m(\kappa L)$ is independent of the contact angle, $\theta_1$, and the cylinder radius, $R$. The dependence of $\Delta h$ on $R$ and $\theta_1$ enters through $\Delta h (\textrm{elliptic})$ and the $\kappa R (1+\sin \theta_1)^{0.12}$ term in Eq.~(\ref{eq:height_final}).

\begin{figure}[ht]
  \centering
  \includegraphics[width = 0.45 \textwidth]{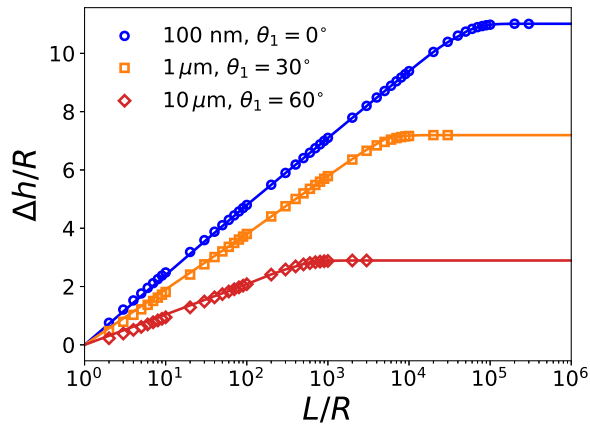}
    \caption{The numerical solutions of the meniscus height (symbols) at various combinations of $R$ and $\theta_1$ as a function of $L$ are compared to the analytical expression in Eq.~(\ref{eq:height_final}).}
	\label{fg:fig5}
\end{figure} 

In Fig.~\ref{fg:fig5}, the analytical result on the meniscus height $\Delta h$ in Eq.~(\ref{eq:height_final}) is compared with numerical solutions of the full Young-Laplace equation rewritten as Eq.~(\ref{eq:YL_1st_order}). A very good agreement has been found between the two, indicating that Eq.~(\ref{eq:height_final}) can be used to accurately predict the meniscus height on the outside of a circular cylinder with $R \ll \kappa^{-1}$ for a meniscus with an arbitrary lateral span, including the crossover zone $0.4\kappa^{-1} \lesssim L \lesssim 4\kappa^{-1}$. However, Eq.~(\ref{eq:height_final}) is a result obtained by comparing the analytical expression of the meniscus height in the limit of $\kappa L \ll 1$ and the numerical results for a full range of $\kappa L$. It remains an open question if the universal expression of $\Delta h$ in Eq.~(\ref{eq:height_final}) for arbitrary $L$, $R$ (as long as it is less than $\kappa^{-1}$), and $\theta_1$ can be derived with an analytical approach.

The results presented in Figs.~\ref{fg:fig4} and \ref{fg:fig5} are for $L$ changing from $2R$ to a value much larger than $\kappa^{-1}$ and for $R$ changing from 100 nm to 10 $\mu$m, i.e., $R \ll \kappa^{-1}$. There are several limits that are of interest but not explored in detail in this paper. In one, if $R$ is made much larger than $\kappa^{-1}$, then the system depicted in Fig.~\ref{fg:fig1} can be regarded as a meniscus between two flat walls (or even be reduced to a meniscus on one flat wall if $L-R \gg \kappa^{-1}$) \cite{Landau1987fluid}. There is a crossover from the $R \ll \kappa^{-1}$ limit, the focus of this paper, and the $R\gg \kappa^{-1}$ limit. In the crossover, $R$ is comparable to $\kappa^{-1}$ and the numerical procedure of dealing with the Young-Laplace equation rewritten as Eq.~(\ref{eq:YL_1st_order}) can be applied. In the limit of $R$ being reduced to nanometer scales, the line-tension effect associated with the large curvature ($R^{-1}$) of the contact line on the surface of the cylinder may become important \cite{Law2017}. In the case where $L-R$ is small enough, factors including disjoining pressure will kick in \cite{Cheng2014PRE}. If $L-R$ is further reduced such that the molecular nature of a liquid has to be taken into account, the continuum theory of capillarity may break down \cite{Cheng2014PRE}. These limits are intriguing directions for future studies.

\section{Conclusions}

The problem of a small circular cylinder immersed in a liquid bath has been studied for many years. The focus was mainly on the limit where the liquid bath is much larger than the capillary length (i.e., $L \gg \kappa^{-1} \gg R$) \cite{Derjaguin1946, White1965, Huh1969, James1974, Lo1983, Alimov2014} or on the case where gravity is negligible and the liquid-vapor interface can be described as a catenary (i.e., $\kappa^{-1} \gg L \gg R$) \cite{DeGennes2004,DeBaubigny2015}. In this paper, we provide a comprehensive discussion of the meniscus on the outside of a circular cylinder with $R \ll \kappa^{-1}$ vertically positioned in a liquid bath with lateral span $L$ ranging from microscopic to macroscopic scales. We obtain an analytical solution of the meniscus profile based on elliptic integrals when $\kappa^{-1} \gg L > R$ and the solution reduces to a catenary when $\kappa^{-1} \gg L \gg R$. In these solutions, the meniscus height $\Delta h \sim R\ln (L/R)$. Our numerical solutions of the full Young-Laplace equation for an arbitrary $L$ indicate that $\Delta h$ indeed scales with $R\ln (L/R)$ up to about $L \lesssim 0.4\kappa^{-1}$. In the opposite limit where $L \gtrsim 4\kappa^{-1}$, the meniscus height agrees well with the prediction of the Derjaguin-James formula and scales with $R\ln (\kappa^{-1}/R)$. The range $0.4\kappa^{-1} \lesssim L \lesssim 4\kappa^{-1}$ is the crossover region where the actual value of $\Delta h$ deviates from the prediction of either the analytical solution based on elliptic integrals or the Derjaguin-James formula.

Our analyses reveal a universal behavior that the analytical solution [Eq.~(\ref{eq:height_elliptic})], which predicts $\Delta h \sim R\ln (L/R)$, always reaches the upper bound set by the Derjaguin-James formula at $L \approx 1.85\kappa^{-1}$. Therefore, the analytical solution with its parameter $l=L/R$ when $L \le 1.85\kappa^{-1}$ and capped at $l=1.85\kappa^{-1}/R$ when $L > 1.85\kappa^{-1}$ can be used to estimate the meniscus height $\Delta h$. The relative deviation of the actual value of $\Delta h$ determined via numerical solutions from this estimate is found to be only noticeable in the crossover region but still less than 5\%. We further find that the relative errors at different $R$ and contact angles at the surface of the cylinder, if properly scaled, as a function of $\kappa L$ all collapse to a master curve. With a fitting function to this master curve, we obtain an analytical expression [Eq.~(\ref{eq:height_final})] that can be used for accurate prediction of $\Delta h$ for the whole range of $L$ from microscopic to macroscopic scales including the crossover zone. Although in this paper we only consider cases with the contact angle on the wall being fixed at $\pi/2$, the theoretical analyses and numerical treatments of the general Young-Laplace equation can also be extended to more general cases with other contact angles at the wall surface. 

\section*{Acknowledgements}
Acknowledgement is made to the Donors of the American Chemical Society Petroleum Research Fund (PRF \#56103-DNI6), for support of this research. S.C. thanks the partial support from a 4-VA Collaborative Research Grant (``Short-range interactions of interfacial bubbles'').

\appendix

\section{Derivation of Young-Laplace Equation} \label{sec:app1}

The profile of a meniscus is governed by Eq. (\ref{eq:Young-Laplace}), which has been discussed extensively for the geometry of sessile and pendant drops. Here we provide a simple derivation of this equation. The energy of a liquid bath bound by a cylindrical container and a meniscus on the outside of a cylinder at the center of the container (Fig.~\ref{fg:fig1}) is a sum of surface energy and gravitational terms, $G = \gamma S + \Delta p V + U_g$, where $\gamma$ is the surface tension of the liquid, $S$ is the surface area of the liquid-vapor interface, $\Delta p$ is a Lagrange multiplier, $V$ is the volume of the liquid bath which is fixed, and $U_g$ is the potential energy of the liquid in the gravitational field. The meniscus profile can be found by minimizing $G$, which can be written in terms of the surface profile $z(r)$,
\begin{align} \label{eq:derivation}
G &=& 2 \pi \gamma \int_{R}^{L} r \sqrt{1 + z'^2} \, \mathrm{d}r + 
2 \pi \Delta p \int_{R}^{L} rz \, \mathrm{d}r \nonumber \\
& & +  \pi \Delta \rho g \int_{R}^{L} r z^2 \, \mathrm{d}r~,  
\end{align}
We seek the surface profile that will make the energy function $G = \int f(z, z', r) \, \mathrm{d}r$ stationary, i.e., $\delta G = 0$. The resulting Euler-Lagrange equation is 
\begin{align}
\frac{\mathrm{d}}{\mathrm{d}r}\frac{\partial f}{\partial z'} 
- \frac{\partial f}{\partial z} = 0~.
\end{align}
After some algebra, we obtain the following equation, 
\begin{align}
\gamma \Big[ \frac{z''}{(1 + z'^2)^{3/2}} + \frac{z'}{r (1 + z'^2)^{1/2}} \Big]
= \Delta p + \Delta \rho g z~, 
\end{align}
where the left hand side comes from the surface energy and the right hand side originates from the volume of the liquid bath being fixed and the gravitational potential energy, respectively. This equation is Eq.~(\ref{eq:Young-Laplace}) in the main text.

\section{Solution of Zero-order}\label{sec:app2}

If the contact angle $\theta_1$ on the cylinder in Fig.~\ref{fg:fig1} is close to $\pi/2$, the resulting liquid-vapor interface is almost flat since the contact angle on the wall surface is fixed at $\pi/2$. In this case $z' = \tan \phi \ll 1$ and Eq. (\ref{eq:Young-Laplace}) can be approximated as 
\begin{equation}
2\tilde{H}  + \kappa^2 z = \frac{1}{r} \frac{\mathrm{d}}{\mathrm{d}r} \Big[ 
\frac{r z'}{(1 + z'^2)^{1/2}} \Big] 
\approx \frac{1}{r} \frac{\mathrm{d}}{\mathrm{d}r} \Big[ 
r z' (1 + \mathcal{O}(z'^2) )\Big]~,
\label{eq:0-order}
\end{equation}
with the following boundary conditions,
\begin{subequations}  
  \begin{align}
	\phi &= \phi_1 \quad \text{at} \quad r = R~, \label{eq:bc9a} \\
	\phi &= \phi_2 \quad \text{at} \quad r = L \quad \text{and} \quad z = 0~, \label{eq:bc9b} 
  \end{align}
\end{subequations}
where $\phi_1 = \theta_1 + \pi/2$, $\phi_2 = \pi$. The solution of Eq.~(\ref{eq:0-order}) which satisfies the boundary condition Eq. (\ref{eq:bc9b}) is,
\begin{align}
z = \frac{2\tilde{H}}{\kappa^2} \Big[ \frac{K_0(\kappa r)}{K_0(\kappa L)} - 1 \Big]~,
\end{align}
and the angle $\phi$ is given by
\begin{align}
\tan \phi = -\frac{2\tilde{H}}{\kappa} \frac{K_1(\kappa r)}{K_0(\kappa L)}~,
\end{align}
where $K_0$ and $K_1$ are modified Bessel functions of second kind of order zero and 
one, respectively. The undetermined constant $\tilde{H}$ can be found using the other
boundary condition Eq. (\ref{eq:bc9a}) and the result is
\begin{align}
\tilde{H} = - \frac{\kappa}{2} \tan \phi_1 \frac{K_0 (\kappa L)}{K_1 (\kappa R)}~.
\end{align}

\renewcommand{\thefigure}{D\arabic{figure}}
\setcounter{figure}{0}
\begin{figure}[hb]
  \centering
  \includegraphics[width = 0.45 \textwidth]{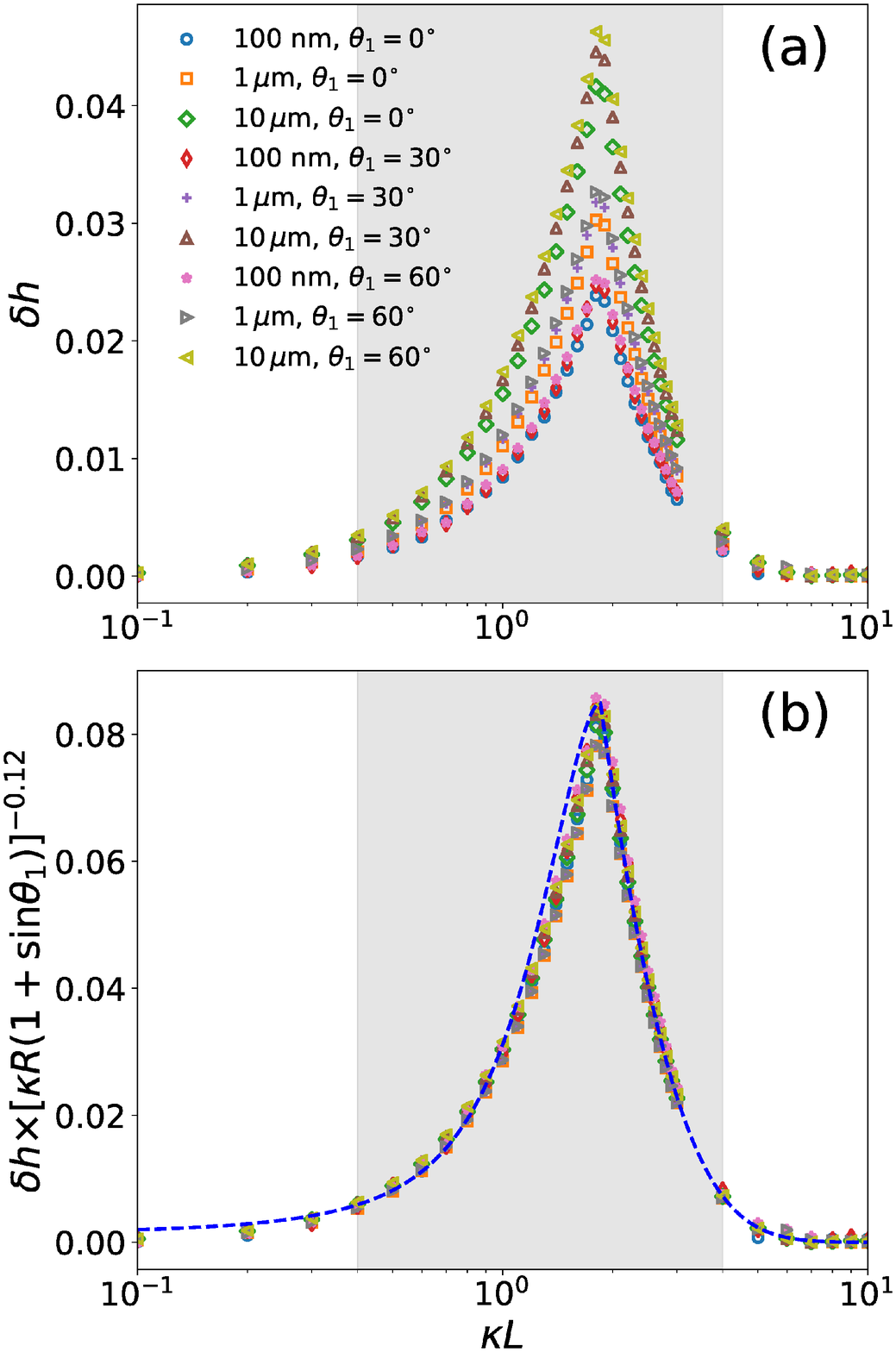}
    \caption{(a) The relative error $\delta h$ defined in Eq.~(\ref{eq:relative_error}) as a function of $\kappa L$ for various combinations of $R$ and $\theta_1$. (b) Data in (a) are collapsed onto a master curve when $\delta h \times [\kappa R (1+\sin \theta_1)]^{-0.12}$ is plotted against $\kappa L$; the blue dashed line is the fit in Eq.~(\ref{eq:fit_kink_function}). In both (a) and (b) the gray zone indicates the crossover region $0.4\kappa^{-1} \lesssim L \lesssim 4\kappa^{-1}$.}
	\label{fg:figS1}
\end{figure} 

\section{Expansion of Elliptic Integrals}\label{sec:app3}

Here we derive the series expansions of incomplete elliptic integrals $F(\phi, j)$ and $E(\phi, j)$ in the limit of $j^2 \rightarrow -\infty$. To facilitate the discussion it is helpful to introduce a small parameter $\epsilon > 0$ and $j^2 = - \frac{1}{\epsilon^2}$; the limit $j^2 \rightarrow -\infty$ thus corresponds to $\epsilon \rightarrow 0$. Below we use the incomplete elliptic integral of second kind, $E(\phi, j)$, as an example. A similar expansion can be performed for $F(\phi, j)$.
\begin{widetext}
\begin{align} \label{eq:E_series}
E(\phi, j) &= \int_{0}^{\phi} \sqrt{1 - j^2 \sin^2 t} \, \mathrm{d}t \quad (t \mapsto \sin t) \nonumber \\
&= \frac{1}{\epsilon} \Big[ \underbrace{\int_{0}^{\sqrt{\epsilon}} \sqrt{\epsilon^2 + t^2} \, \frac{\mathrm{d}t}{\sqrt{1 - t^2}} }_{t \, \mapsto \, \epsilon t}
+ \underbrace{\int_{\sqrt{\epsilon}}^{\sin \phi} \sqrt{\epsilon^2 + t^2} \, \frac{\mathrm{d}t}{\sqrt{1 - t^2}} }_{ t \, \mapsto \, 1/t } \Big] \nonumber \\
&= \frac{1}{\epsilon} \Big[ \epsilon^2 \int_{0}^{1/\sqrt{\epsilon}} \sqrt{1 + t^2} \, \frac{\mathrm{d}t}{\sqrt{1 - \epsilon^2 t^2}} + \int_{1/\sin \phi}^{1/\sqrt{\epsilon}} \sqrt{1 + \epsilon^2 t^2} \, \frac{\mathrm{d}t}{t^2 \sqrt{t^2 - 1}} \Big]  \nonumber \\
& = \frac{1}{\epsilon} (1 - \cos \phi) + \epsilon (-\frac{\ln \epsilon}{2} + \ln 2 + \frac{1}{4} - \frac{1}{2} \ln \frac{1 + \cos \phi}{\sin \phi}) + \mathcal{O} (\epsilon^2) ~.
\end{align}
\end{widetext}
In this derivation we have employed the following expansion $\frac{1}{\sqrt{1 - \epsilon^2 t^2}} = 1 + \frac{1}{2}\epsilon^2 t^2 + \mathcal{O}(\epsilon^4 t^4)$ and $\sqrt{1 + \epsilon^2 t^2} = 1 + \frac{1}{2} \epsilon^2 t^2 + \mathcal{O}(\epsilon^4 t^4)$, and assumed that $\sin \phi > \sqrt{\epsilon}$. The expansion of the incomplete elliptic integral of first kind, $F(\phi, j)$, can be obtained similarly and the result is
\begin{align} \label{eq:F_series}
F(\phi, j) &= \epsilon (-\ln \epsilon + 2 \ln 2 - \ln \frac{1 + \cos \phi}{\sin \phi}) + \mathcal{O}(\epsilon^2)~.
\end{align}
By substituting Eq.~(\ref{eq:E_series}) and Eq.~(\ref{eq:F_series}) into Eq.~(\ref{eq:height_elliptic}), we arrive at 
\begin{align}
\Delta h = R \cos \theta_1 \Big[\ln  \frac{2L}{R(1 + \sin \theta_1)} -\frac{1}{2}\Big]~,
\end{align}
which is Eq.~(\ref{eq:height_logL_2}) in the main text. Here the relations $H = \frac{\sin \phi_1}{l^2 - 1} \approx \frac{\cos \theta_1}{l^2}$ and $\epsilon = \sqrt{-j^{-2}}= \sqrt{c}$ are used.

\section{Relative Error of Eq.~(\ref{eq:height_elliptic}) on Predicting $\Delta h$}\label{sec:app4}

In order to obtain an even more accurate expression of the meniscus height that applies to $R \ll \kappa^{-1}$ and an arbitrary $L$, we denote the meniscus height predicted in Eq.~(\ref{eq:height_elliptic}) using elliptic integrals with the parameter $l$ given in Eq.~(\ref{eq:height_all}) as $\Delta h (\textrm{elliptic})$. The full numerical solution of Eq.~(\ref{eq:YL_1st_order}) for an arbitrary $L$ is denoted as $\Delta h (\textrm{actual})$. The relative error of using Eq.~(\ref{eq:height_elliptic}) to predict the meniscus height is thus given by
\begin{equation}
\delta h = \frac{\Delta h (\textrm{elliptic}) - \Delta h (\textrm{actual})}{\Delta h (\textrm{elliptic})}~.
\label{eq:relative_error}
\end{equation}
In Fig.~\ref{fg:figS1}(a), $\delta h$ is shown as a function of $L$ that is normalized by $\kappa^{-1}$ for several combinations of the cylinder radius, $R$, and the contact angle on its surface, $\theta_1$. As expected, the peak value of the relative error occurs at $\kappa L = 1.85$. We find that all the data collapse to a master curve if we plot $\delta h \times [\kappa R (1+\sin \theta_1)]^{-0.12}$ against $\kappa L$, as shown in Fig.~\ref{fg:figS1}(b). The master curve can be fit with the kink function given in Eq.~(\ref{eq:fit_kink_function}) [dashed blue line in Fig.~\ref{fg:figS1}(b)]. With this universal fit to the collapsed data of relative error, we arrive at Eq.~(\ref{eq:height_final}) in the main text that can be used to accurately predict the meniscus height for an arbitrary $L$.

%\bibliography{cylinder}


\end{document}